\documentclass[conference]{IEEEtran}

\usepackage{algorithm, algorithmic}
\usepackage{amsmath, amssymb}
\usepackage{graphics, graphicx}
\usepackage{color, xcolor, soul}
\usepackage{booktabs}
\usepackage{multirow}
\usepackage{newcent}

\colorlet{shadecolor}{yellow}

\begin{document}

\title{A Techno-economic Analysis of Network Migration to Software-Defined Networking}

\author{\IEEEauthorblockN{Tamal Das, Marcel Caria and Admela Jukan}
\IEEEauthorblockA{Technische Universit\"at Carolo-Wilhelmina zu Braunschweig\\
Email: \{das, caria, jukan\} @ida.ing.tu-bs.de}
\and
\IEEEauthorblockN{Marco Hoffmann}
\IEEEauthorblockA{Nokia Siemens Networks\\
Email: marco.hoffmann@nsn.com}
}

\maketitle

\begin{abstract}
As the Software-Defined Networking (SDN) paradigm gains momentum, every network
operator faces the obvious dilemma -- when and how to migrate from existing IP
routers to SDN-compliant equipments. A single-step
complete overhaul of a fully functional network is impractical, while at the
same time, the immediate benefits of SDN are obvious. A viable solution is thus
a \emph{gradual} migration over time, where questions of which routers should
migrate first, and whether the order of migration makes a difference, can be
analyzed from techno-economic and traffic engineering perspective. In this
paper, we address these questions from the techno-economic perspective, and
establish the importance of \emph{migration scheduling}. We propose optimization
techniques and greedy algorithms to plan an effective migration schedule, based
on various techno-economic aspects, such as technological gains in combinations
with CapEx limitations. We demonstrate the importance of an effective migration
sequence through two relevant network management metrics, namely, number of
alternative paths availed by a node on migration, and network capacity savings.
Our results suggest that the sequence
of migration plays a vital role, especially in
the early stages of network migration to SDN.
\end{abstract}

\section{Introduction}\label{sec:intro}
Software Defined Networking (SDN) is an emerging networking architecture
paradigm that separates the control plane from the data plane, facilitating
programmability of the network control functions \cite{SDN-onf-whitepaper}. SDN
is expected to reduce the network OpEx by simplifying operations, optimizing
resource usage through centralized data/algorithms, and simplifying network
software upgrades. It also significantly cuts down a network operator's CapEx,
since a commercial-off-the-shelf (COTS) server with a high-end CPU is much
cheaper than a high-end router \cite{metaswitch-whitepaper}.
As the SDN paradigm gains momentum, the migration from existing IP routers to
SDN-compliant equipment, e.g., OpenFlow switches, is becoming eminent. In data
centers, SDN can be already fully integrated into the network architecture, by
migrating the switching and routing infrastructure entirely to SDN. For a
medium- to large-scale ISP, on the other hand, a viable approach is to gradually
migrate to SDN, for instance, over a multi-period cycle spanning couple of
years.

\par In cases where due to operational and economic considerations not all
routers in a fully functional network can be replaced at once, the operator
would need to know which network nodes should be migrated first, and which ones
later. Some operators may choose to migrate a fixed number of routers in their
network, say quarterly, until complete migration of the network. Others may have
a hard limit on their CapEx investment available for migration every quarter,
and may choose to quarterly migrate variable number of routers within their
CapEx limit.
In either case, the operators need to understand how their network can make best
use of traffic engineering capabilities enabled by SDN, during various stages of
a migration process. Especially those stages are of interest, wherein the native
IP routing, such as OSPF, co-exists with SDN-enabled traffic-engineered routing.

\par In this paper, we propose a techno-economic analysis of network migration
to software-defined networking, which includes aspects of migration scheduling.
To this end, we design the corresponding optimization model, and propose
effective greedy algorithms for the same. Although a scheduled migration must
intuitively be better than a random one, as our study confirms, our approach
exhibits low computational complexity and high degree of optimality, which makes
it highly practical. We consider two scenarios of migration, one where an
Internet Service Provider (ISP) has a limit on the number of routers migrated
per time-step, and the other where it has a limit on the amount of CapEx
investment required for migration per time-step. Thus, our main contributions in
this paper are two-fold - (a) simultaneous consideration of technological gains
and economic viability of migration to SDN, and (b) design of novel greedy
algorithms that come remarkably close to the optimal, measured in terms of
traffic engineered metrics, such as number of alternative routes, as well as
network capacity savings.

\par The rest of this paper is organized as follows. Section
\ref{sec:related-work-contri} discusses the related work.
Section \ref{sec:reference-scenario} presents the reference migration scenario.
In Section \ref{sec:migration-model}, we present our migration models, discuss
their complexity and optimality, and evaluate the same using simulations in
Section \ref{sec:results}.
Section \ref{sec:conclusion} concludes our paper.

\section{Related work}\label{sec:related-work-contri}
Network migration has been studied using \emph{system-dynamics} and
\emph{agent-based} models. In the system dynamics approach, the migration
problem is treated as a dynamic system \cite{Sen10}, where the rate of
migration depends on number of migrated agents in the system. On
the other hand, in an agent-based approach \cite{Macy02}, the system consists of
an ensemble of agents, each trying to increase its own utility. Such studies
have mostly been conducted for IPv6 \cite{Trinh10} and secure BGP \cite{Gill11}.

\par A choice of the prediction interval for the demand and cost forecast has
been studied in \cite{Verbrugge}, where it was shown that short prediction
intervals may not be able to sufficiently account for future evolutions, while
long prediction intervals may be uncertain. An efficient ant colony meta
heuristic for the multi-layer, multi-period migration problem formulated as a
path-finding problem is proposed in~\cite{antColony}. Stochastic programming
approaches have also been proposed to better handle the uncertainties, as in
\cite{Schupke}.
Timing issues of migration to a new technology have also been studied, and it
was shown that demand growth, migration cost, and cost savings from the new
technology must be considered \cite{Rajagopalan}. In~\cite{McKeown}, the number
and location of SDN controllers is studied, which is an important related
aspect, but outside the scope of this paper.

\begin{figure}[t]
\begin{center}
\includegraphics[width=8.6cm]{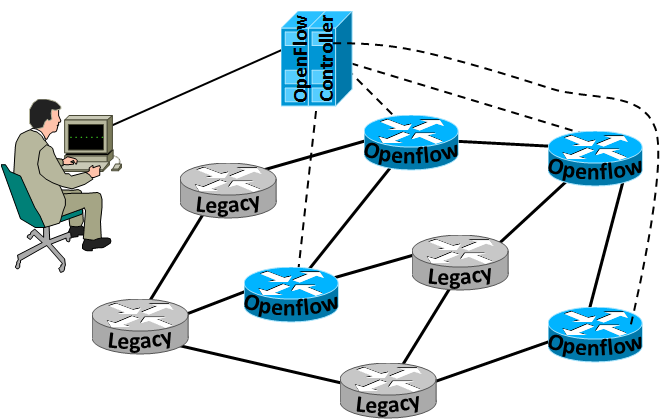}
\caption{Reference migration scenario}
\label{fig:reference-migration-scenario}
\end{center}
\vspace{-5mm}
\end{figure}

\par This paper extends our recent studies on network migration.
In \cite{netmig-icc2012}, we proposed an agent-based model to study benefits of
joint migration to multiple technologies, on a case study of Path Computation
Element (PCE) and SDN. In \cite{caria-sdn-migration}, we defined the SDN
\emph{migration scheduling problem}, and found optimal solution.
In this paper, however, our focus is entirely on devising computationally
inexpensive but effective heuristics to schedule migration to a single
technology, with consideration of techno-economic factors, such as limitations
on CapEx investment. To this end, we propose novel greedy algorithms and
establish their practical impact in terms of computational complexity, run
times, network capacity gains, etc.

\section{Reference Migration Scenario}\label{sec:reference-scenario}

\par The reference migration scenario is illustrated in Figure
\ref{fig:reference-migration-scenario}, wherein an operator manages a network of
4 nodes each of legacy routers and SDN-compliant routers using a network
management system (NMS). The network is in an intermediate stage of migration,
thereby comprising of both legacy and SDN routers. Whereas legacy routers run
OSPF, the SDN routers are collectively controlled by a centralized OpenFlow
controller, which is in turn managed by the NMS in providing the TE capabilities
in the sense of network optimization. Here, what SDN does provide is a thorough
configurability of the routing of all flows in the network, and thus a much
greater solution space for network optimization, which in turn leads to an
overall better performance of TE. Furthermore, we assume that the SDN routers
are also OSPF-compliant, in that they are capable of exchanging OSPF packets
with legacy routers. This feature is fundamental to gradual migration and can
either be implemented in the SDN routers, or the SDN controller
\cite{routeflow}.

\par To facilitate traffic engineering actions through SDN, we assume the SDN
controller to have access to all necessary information including network
topology, traffic monitoring, and routing. Without loss of generality, we assume that
traffic engineering is performed only once at the beginning of every migration
period. Therefore, a reasonable capacity headroom (e.g., maximum link
utilization set to 70\%) assures that despite the expected traffic growth, all
links are loaded fairly below the threshold until the end of the current
migration period. Before the migration starts, routing is based on the OSPF
protocol only, i.e., traffic is always forwarded on the shortest path based on
the destination IP address. Finally, traffic engineering through OSPF link cost
(weight) changes is not considered in this paper, which we justify with the
known fact that network operators do not commonly deploy OSPF link weights for
the reasons of routing stability~\cite{dynamicRouting}. 
% For instance, Cisco
% recommends to just set the link cost inversely proportional to the capacity of
% the link, without taking its utilization into account~\cite{linkCost}.

\par It is important to note that our focus in this paper is on IP layer, and
not on network technologies with built-in traffic engineering capabilities,
e.g., MPLS-TE. The reason behind that is, as we believe, SDN routers will
substitute legacy IP routers, whereas other technologies work in the layers
below, and do not allow gradual migration; for instance, one cannot deploy a few
Carrier Ethernet switches in combination with legacy SONET equipment.

% \par We also account for periodic growth in traffic, which is usually
% accommodated by network operators by means of link capacity upgrades. However,
% our goal is to investigate whether advanced traffic engineering capabilities
% of SDN-enabled routers can reduce the need for such upgrades.

\section{Optimizations and Greedy Algorithms}\label{sec:migration-model}
In this section, we present different optimization techniques and greedy
algorithms for SDN migration. We consider two scenarios, one where
the ISP has a limit on the number of routers migrated per
time-step, and the other where it has a limit on the amount of CapEx investment
required for migration per time-step. We first summarize the overall migration
scheduling strategy from our previous work \cite{caria-sdn-migration}, and then
discuss our novel contributions here. 

\par
\begin{figure}[t]
\begin{center}
\includegraphics[width=7cm]{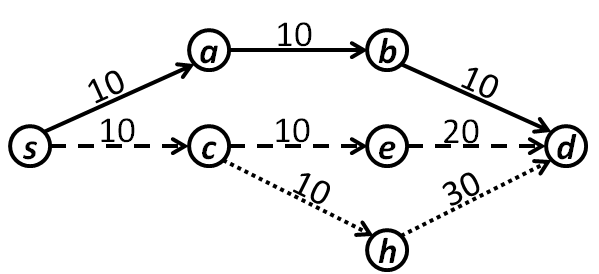}
\caption{Illustration of key-nodes of a path}
\label{fig:key-nodes}
\end{center}
\vspace{-5mm}
\end{figure}

 To migrate a network, for every node pair, we first compute the least-cost
path, as well as other paths between the same node pair with hop-length equal to
its least-cost path. These are the paths between a node pair that become
available on migration to SDN, thereby enabling traffic engineering in the
network. We then identify the \emph{key-nodes} on each path, which are those
nodes that \emph{must} be SDN-compliant for a path to be used for traffic
engineering. Thus, for the least cost path, there are no key nodes, and
migration of all key-nodes ensures availability of all possible paths in the
network for traffic engineering. The concept of key-node is illustrated using
Figure \ref{fig:key-nodes}, which shows the least-cost path between $s$ and $d$,
as well as two more paths of same hop-length as that of the least-cost path. The
link weights are indicated above each link. As defined, there are no key-nodes
for the least-cost path $s-a-b-d$.
For the path $s-c-e-d$ to be available, $s$ must be SDN-compliant, and thus $s$
is its key-node. Finally, for the path $s-c-h-d$ to be
available, both $s$ and $c$ must be SDN-compliant, and thus those are its
key-nodes.

\par The generic approach to compute these key-nodes for a path is summarized in
Algorithm \ref{algo:key-nodes}, which essentially involves computation of the
\emph{fork} node between two paths between the same node pair. A path is thus
available only after \emph{all} its key-nodes migrate. The subsequent step is thus to choose the
sequence of key-nodes for migration, which we attempt using optimization and
greedy approaches. 

\begin{algorithm}[b]
\caption{Key-node computation for a path} \label{algo:key-nodes}
\begin{algorithmic}
\STATE {\textbf{Input:} Path $p$}
\STATE {\textbf{Output:} $K_p \gets$ Set of key-nodes of path $p$}
\STATE {$K_p \gets \phi$}
\STATE {$A \gets$ Set of paths of cost lesser than that of $a$, between source
and destination of $p$}
\FORALL {$a \in A$}
	\STATE {$K_p \gets K_p \cup$ \{last common node between $a$ and $p$
	from source to destination of $p$\}}
\ENDFOR
\RETURN {$K_p$}
\end{algorithmic}
\end{algorithm}

\subsection{Optimization Approaches} \label{sec:ILP}
We now present Integer Linear Programs (ILP) to determine the optimal sequence
of migration of nodes in a network, with and without cost constraints.
We base our ILP on the ILP formulation from \cite{caria-sdn-migration}, both for
comparison and completeness, and further extend it to include CapEx constraints.
The parameters and variables used in our model are summarized in Table
\ref{table:symbols}. Here, we attempt to maximize the cumulative number of paths
available for traffic engineering during the course of migration as our
objective, i.e., $$ \text{Maximize\,\,\,} \sum_t\sum_p \pi^p_t \cdot \varphi_p
$$

\par The availability of path $p$ in time-step $t$ depends on whether all of
its key-nodes ($\alpha_p$ in number) have migrated till that time-step. Thus,
$$
\forall p, t: \,\,\,\,\,\,\,\,\,\,
\pi^p_t \cdot \alpha_p \leq \sum_n \beta^n_p \cdot \mu^n_t
$$

\par In the case without CapEx restrictions, the total number of nodes that may
migrate in a time-step is bounded by the total number of nodes in the network
averaged over the entire migration period. Whereas, in the case with CapEx
restriction on each time-step, the average CapEx per time-step invested till
time-step $t$ must not exceed the average CapEx per time-step over the entire
migration period.
Thus, 
$$
\begin{array}{c}
\forall t: \sum_n \mu^n_t \leq t \left\lceil    \dfrac{|N|}{T}  \right\rceil\\
(\text{without cost})\\
\end{array}
\,\,\,\,\,\,\text{or}\,\,\,\,\,\,
\begin{array}{c}
\forall t:
\sum_n \mu^n_t \cdot c_n \leq t \left(   
\dfrac{C}{T} \right) \\
(\text{with cost})\\
\end{array}
$$

Finally, we restrict a node that has once migrated from \emph{reverse-migrating}, i.e.,
\[
\forall t: \,\,\,\,\,\,\,\,\,\,
\mu^n_t \leq \mu^n_{t+1}
\]

\begin{table}[tb]\begin{center}%\scriptsize
\caption{List of Notations Used}\label{table:symbols}
\begin{tabular}{ c l }
\toprule
\textbf{Parameter} & \multicolumn{1}{c}{\textbf{Meaning}} \\
\midrule
$0\leq t\leq\text{T}$ & Migration time-step $t$, number of time-steps T
\\\addlinespace[1.0mm] $p\in\text{P}$ & Path $p$, set of all pre-calculated paths P \\\addlinespace[1.0mm]
$n\in\text{N}$ & Network node $n$, set of all nodes N \\\addlinespace[1.0mm]
$\varphi_p$ & Priority of path $p$ \\\addlinespace[1.0mm]
$c_n$ & Cost of migrating node $n$ \\\addlinespace[1.0mm]
C & Total CapEx required to migrate all nodes \\\addlinespace[1.0mm]
{$\alpha_p$} & Number of key nodes on path $p$ \\\addlinespace[1.0mm]
\multirow{2}{*}{$\beta^n_p$} & Boolean routing parameter, true \\
 & $\text{ }$ if node $n$ is a key node on path $p$ \\\addlinespace[1.0mm]
\midrule
\textbf{Variable} & \multicolumn{1}{c}{\textbf{Meaning}} \\
\midrule
\multirow{2}{*}{$\mu^n_t$} & Boolean to determine if node $n$ is \\
 & $\text{ }$ migrated till time-step $t$ \\\addlinespace[1.0mm]
\multirow{2}{*}{$\pi^p_t$} & Boolean to determine if path $p$ is \\
 & $\text{ }$ available in time-step $t$ \\\addlinespace[1.0mm]
\bottomrule
\end{tabular}
\vspace{-5mm}
\end{center}\end{table}

\subsection{Greedy Algorithms} \label{sec:greedy-algo}
In our greedy approach to determine a suitable migration schedule of nodes in a
network, we essentially focus on the number of additional alternative paths
availed by a node on migration. This is summarized in Algorithms
\ref{algo:greedy-algo},
\ref{algo:node-selection-without-cost} and \ref{algo:node-selection-with-cost}.
A preliminary operation to our greedy approach is a one-time computation of 
least-cost path for each node pair, as well as the paths between each node
pair equal in hop-length to the corresponding least-cost path.

\par At each time-step, we compute the list of all unmigrated key-nodes in the
network, and then compute the number of additional alternative paths that can be
availed by migration of each of these key-nodes. For example, in Figure
\ref{fig:key-nodes}, if node $c$ migrates, then the path $c-h-e$ will be an
additional alternative path available to $c$ for traffic engineering. In the
case without CapEx restrictions, at each time-step, we sort the nodes in
descending order of the number of alternative paths made available by their
migration, and select the first $\lceil N/T \rceil$ nodes for migration.
In the case with CapEx limitations on each time-step, we first compute the total
CapEx available at current time-step, which is sum of the average CapEx per
time-step, and the residual CapEx from all previous time-steps. The residual
CapEx results from partial utilization of the CapEx available for a time-step.
We then sort the unmigrated key-nodes in decreasing order of the ratio of the
number of alternative paths availed by migration of a node and the corresponding
cost of migration, and select as many nodes for migration as possible while
traversing this sorted list as is allowed by the available CapEx for the current
time-step.
Algorithm \ref{algo:node-selection-with-cost} thus selects nodes for migration
based on the number of additional paths per unit cost of migration.

\par For example, say for a network, $m$ = 2, and the cost of migration of nodes
key-nodes $a$, $b$, and $c$ are 5, 8 and 10 units, respectively. If
at a particular time-step, $a$, $b$, and $c$ can respectively avail of 2, 4 and
3 additional alternative paths on migration, and the CapEx available for this
time-step is 15 units, then as per Algorithm
\ref{algo:node-selection-without-cost}, $b$ and $c$ are selected for migration,
whereas, as per Algorithm \ref{algo:node-selection-with-cost}, $a$ and $b$
should migrate.

\begin{algorithm}
\caption{Greedy algorithm for migration schedule} \label{algo:greedy-algo}
\begin{algorithmic}
\STATE {\textbf{Input:} $T \gets$ Number of time-steps}
\FOR{$t \gets 1$ to $T$}
	\STATE {$U \gets$ List of unmigrated key-nodes in the network}
	\FORALL {$u \in U$}
		\STATE {$P(u) \gets$ Number of additional alternative paths available by
		migration of node $u$}
	\ENDFOR
	\STATE {Based on $P$, use \textbf{Algorithm 2} or \textbf{3} to select the
	node(s) to migrate}
\ENDFOR
\end{algorithmic}
\end{algorithm}

\begin{algorithm}
\caption{Node selection algorithm without costs}
\label{algo:node-selection-without-cost}
\begin{algorithmic}
\STATE {\textbf{Input:} $P$}
\STATE {$\,\,\,\,\,\,\,\,\,\,\,\,\,\,\,\,m \gets$ Number of nodes to
migrate per time-step}
\STATE {Sort $P$ in descending order}
\FOR {$i \gets 1$ to $m$}
	\STATE {Migrate node $i$}
\ENDFOR
\end{algorithmic}
\end{algorithm}

\begin{algorithm}
\caption{Node selection algorithm based on costs}
\label{algo:node-selection-with-cost}
\begin{algorithmic}
\STATE {\textbf{Input:} $P$}
\STATE {$\,\,\,\,\,\,\,\,\,\,\,\,\,\,\,\,\,\,\,\,\,C_t \gets$ CapEx available
for time-step $t$}
\STATE {$\,\,\,\,\,\,\,\,\,\,\,\,\,\,\,\,\,\,\,\,\,c_i \gets$ cost of migrating
node $i$}
\FOR {$i \gets 1$ to $|P|$}
	\STATE {$B(i) \gets \dfrac{P_i}{c_i}$}
\ENDFOR
\STATE {Sort $P$ in descending order of corresponding entries in $B$}
\FOR {$i \gets 1$ to $|P|$}
	\IF {$C_t \leq c_i$}
		\STATE {$C_t \gets C_t - c_i$}
		\STATE {Migrate node $i$}
	\ELSE
		\STATE {Break out of \textbf{for} loop}
	\ENDIF
\ENDFOR
\end{algorithmic}
\end{algorithm}

\subsection{Computational Complexities}

\par To establish the novelty of our greedy approach over the ILP, we compare
the time complexities of both approaches. The ILP efficiently scans the entire
search space for optimal solutions. 

\par The ILP primarily comprises of a
combinatorial problem of partitioning a set $S$ of $N$ elements into its
subsets $\{S_i: 1 \leq i \leq T\}$, i.e., 
$$ |S| =
\prod_{i=1}^{T-1} \binom{z_i}{|S_i|}, \text{where}, 
$$

$$
z_i=
\left\{\begin{array}{c c}
N & i=1\\
N-\sum_{j=1}^{i-1}|S_j| & 1 < i < T \\
\end{array} \right.
$$

We consider all except the last subset to be of equal size $\lceil n/T
\rceil$, thereby resulting in the last subset of size $N -
(T-1)\lceil N/T \rceil$.
Thus, $$
|S_i|=
\left\{\begin{array}{c c}
\lceil N/T  \rceil & 1 \leq i < T\\
N - (T-1)\lceil N/T  \rceil & i=T\\
\end{array} \right.
$$

\par We now derive the time complexity of our greedy algorithm (in section
\ref{sec:greedy-algo}) with and without cost constraints. A preliminary step to
Algorithm \ref{algo:greedy-algo} involves computation of all paths between every
node pair, which can be achieved (say, using Floyd-Warshall Algorithm) in
$O(N^3)$.

\par In Algorithm \ref{algo:greedy-algo}, the outer \texttt{for} loop is of the
order of number of time-steps, i.e. $O(T)$, while the inner \texttt{for} loop
runs across all unmigrated key-nodes, which is $O(N)$. The number of additional
alternative paths made available by migration of a node, can be retrieved by
scanning the list of precomputed paths between every node pair in constant time,
given that we only consider the shortest hop-length paths between two nodes
for the purpose of traffic engineering. The CapEx for a time-step can be computed
based on the CapEx invested in the previous time-steps, resulting in
$O(T)$.

\par In Algorithm \ref{algo:node-selection-without-cost}, $P$ can be sorted in
$O(N \log N)$, whereas, the \texttt{for} loop runs over all nodes, i.e. $O(N)$.
Thus, Algorithm \ref{algo:node-selection-without-cost} results in $O(N+N\log
N)$.

\par In Algorithm \ref{algo:node-selection-with-cost}, the first \texttt{for}
loop runs over the list of unmigrated key-nodes, which is of the order of
$O(N)$. The input element $P$ is of length $O(N)$, and can be sorted in
$O(N \log N)$. The second \texttt{for} loop also runs over the list of
unmigrated key-nodes, resulting in $O(N)$ complexity. Thus, the order
complexity of Algorithm \ref{algo:node-selection-with-cost} is $O(N + N \log
N)$.

\par Cumulatively, the time complexity of Algorithm \ref{algo:greedy-algo} can
be approximated to $O(T\cdot N \log N)$, which comes in sharp contrast with
the exponential complexity of the optimization algorithm.

\section{Simulation and Results}\label{sec:results}
\par For the performance evaluation, we developed a Java-based simulation
framework. The topology used was the TA2 network (65 nodes, 108 links) from the
SNDlib topology library \cite{sndlib}.
We studied the migration profile of the nodes over 10 equal migration periods. We
assigned uniformly random link weights assuming that the difference between any
two link costs is below the inverse of the global maximum path length relative
to the minimum link cost in the topology. This mimics OSPF's behavior in that
between a node pair, no $n$-hop path should cost more than a $(n+1)$-hop path.

The migration sequence optimization was computed with the GUROBI optimizer on an
Intel Core i7-3930K CPU (6 x 3.2 GHz) in less than 10 seconds, while the optimal
traffic distribution computation took about 30 minutes (with an allowed MIP
gap of 1\%). Each of the plots in Figure \ref{fig:capacity
savings} were averaged over 10 different random assignments of link weights and
traffic profiles. Each traffic flow was uniformly distributed between 0 and 400
Mbit/s. We assumed links available in the following granularities (in Gbit/s):
1, 5, 10, 40, 100, 400, and 1000. To incorporate traffic growth during the
course of the simulation, we set the growth factor of each flow in each
time-step to a random value between 1.05 and 1.3, to ensure mean traffic growth of 20\% in
every time-step. We next demonstrate the importance of effective migration
sequence through two relevant metrics, namely, number of alternative available
paths between a node pair, and network capacity savings.

\begin{figure}
\begin{center}
\includegraphics[width=9.5cm]{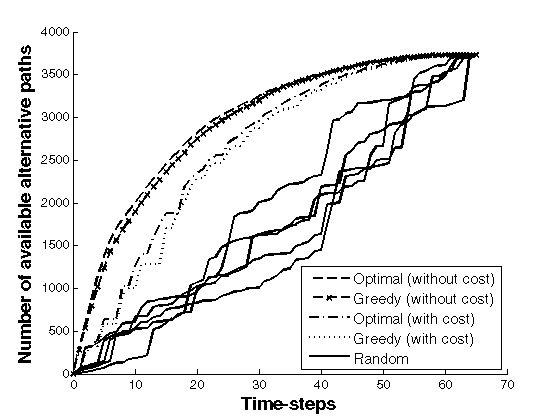}
\caption{Performance comparison of optimal, greedy, and random migration
schedules without cost considerations}
\label{fig:number of migrated nodes}
\end{center}
\vspace{-7mm}
\end{figure}

\par Figure \ref{fig:number of migrated nodes} plots the number of additional
alternative paths resulting over time from various migration sequences -
optimal, greedy, random. At the start of the simulation, none of the nodes had
migrated, resulting in all curves starting at origin.
Similarly, at the end of the simulation all nodes had migrated, resulting in a
single end-point for all curves. The challenge thus remains in optimizing the
\emph{sequence} of migration, which we found to make a big difference in
the number of alternative paths made available through SDN capabilities at
migrated nodes, especially in the early stages. For the scenario without cost
considerations, we restricted the number of nodes that can migrate in a
time-step, namely, 7 nodes in each of the first 9 time-steps, 2 remaining nodes in the
last time-step. For the case with cost constraints, we restricted the amount of
CapEx investment per time-step. The cost of migrating a node was assumed proportional to its degree of
connectivity in the network.

\par At the start of the simulation, we allot the network operator the total CapEx
required for migration of all nodes in its network. It is remarkable to note how
close the greedy algorithm comes to the optimal, both with and without cost
constraints. The CapEx limit per time-step can also be seen from the step-like
pattern of the curves with cost considerations, in contrast with the smooth
curves without cost considerations. It is also noteworthy that the optimal (or
greedy) sequence without cost constraints is higher than the corresponding
curve with cost constraints. This is because the large number of alternative
paths made available in the early stages results from a heavy CapEx investment
in the early stages, which is checked with CapEx constraints. We also found
similar behavior (not shown here for brevity), as in Figure \ref{fig:number of migrated nodes}, for Cost266
network (37 nodes, 57 links) as a medium-size topology and
France network (25 nodes, 45 links) as a small-size topology
from the SNDlib library \cite{sndlib}. 

\begin{figure}
\begin{center}
\includegraphics[width=9.5cm]{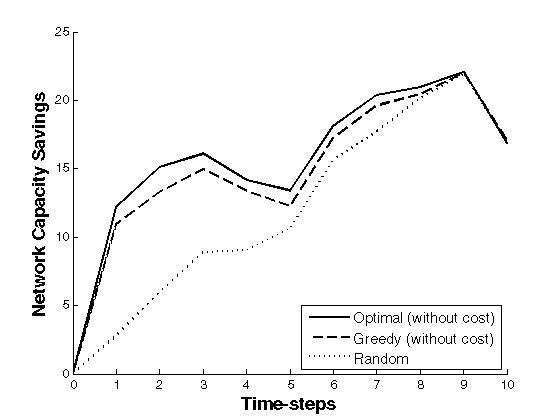}
\caption{Network capacity savings of optimal, greedy, and random migration
schedules}
\label{fig:capacity savings}
\end{center}
\vspace{-5mm}
\end{figure}

\par Figure \ref{fig:capacity savings} plots the network capacity savings over
time resulting from different migration sequences for the case with cost
constraints. We note that the optimal migration sequence significantly
outperforms the random migration sequence, while the greedy approach closely
follows the optimal throughout. The drop observed in the curves at around
the middle and end is due to the first-time use of a higher capacity link, 400
Gbit/s for the drop in the middle, and 1 Tbit/s for the drop at the end. These
links are not fully utilized in the early stages.

\par
\begin{figure}
\begin{center}
\includegraphics[width=9.5cm]{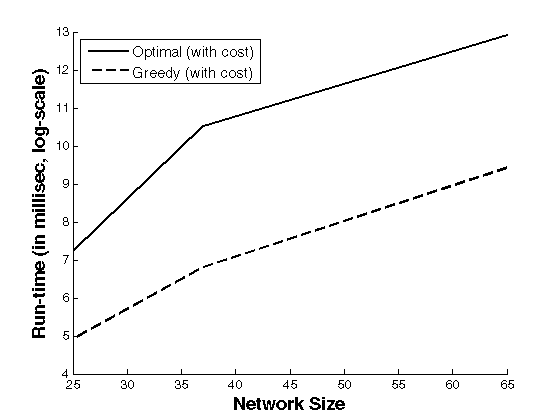}
\caption{Simulation runtime of Optimal and Greedy approach}
\label{fig:run-times}
\end{center}
\vspace{-5mm}
\end{figure}

 In Figure \ref{fig:run-times}, we compare the observed simulation run-times
(in milliseconds, log-scale) of the migration sequence computation for the ILP
and greedy approach for the case with cost constraints.
Each data point in the plot was averaged over 10 runs.
We observe that our heuristic increasingly outperforms the optimization model
with increasing network size. This in turn, establishes the scalability of our
algorithm, when compared to computation of the optimal migration sequence.

\section{Conclusions}\label{sec:conclusion}

\par In this paper, we presented optimization techniques and greedy algorithms to solve for the optimal migration sequence of
SDN routers from the techno-economic perspective, and compared them analytically and empirically. The proposed approach included technological
gains, as well as constraints on migration CapEx investment.
In addition to the significantly lower time complexity, we observed that our
greedy approach comes very close to the optimal, both in terms of number of
alternative paths, as well as network capacity savings.

\bibliographystyle{IEEEtran}
\bibliography{sdn-migration}

\end{document}